# Electrochemical synthesis of iron-based superconductor FeSe films

Satoshi Demura[1,2,3], Toshinori Ozaki[1,3], Hiroyuki Okazaki[1,3], Yoshikazu Mizuguchi[3,4], Hiroshi Hara[1], Yasuna Kawasaki[1,2,3], Keita Deguchi[1,2,3], Tohru Watanabe[1,2], Takahide Yamaguchi[1,3], Hiroyuki Takeya[1,3] and Yoshihiko Takano[1,2,3]

[1]National Institute for Materials Science (NIMS), 1-2-1 Sengen, Tsukuba, Ibaraki 305-0047, Japan

[2]University of Tsukuba, 1-1-1 Tennodai, Tsukuba, Ibaraki 305-8577, Japan

[3]JST-TRIP, 1-2-1 Sengen, Tsukuba, Ibaraki 305-0047, Japan

[4]Tokyo Metropolitan University, 1-1 Minami-Osawa, Hachioji, Tokyo 192-0397, Japan

**Abstract**

The superconducting FeSe films were successfully fabricated using the electrochemical synthesis. The composition ratio of Fe and Se can be controlled by the electric potential and pH value. We found that the FeSe films deposited at the electric potential -1.75 V and pH 2.3 show the superconducting transition at 3.5 K. The establishment of this electrochemical synthesis technique will provide many advantages for application.

Since the discovery of superconductivity in LaFeAsO$_{1-x}$F$_x$[1], several types of iron-based superconductors have been discovered.[2-7] Among them, FeSe with $T_c$ ~8 K is quite promising for practical applications because of the simplest structure and less toxicity compared to other As-based compounds. Furthermore, the superconducting transition temperature $T_c$ dramatically increases up to 37 K under high pressure.[8-11] FeSe films have been fabricated by the physical deposition methods such as PLD and MBE techniques.[12, 13] However, these fabrication processes tend to be expensive, complex, and difficult for large area deposition. We tried to realize the fabrication of FeSe films by electrochemical deposition, since MgB$_2$ films have succeeded by the same method.[14] The electrochemical deposition has further advantages to fabricate FeSe films because of the inexpensive equipment, quick synthesis, and ease of deposition of large area at room temperature. There are some reports about the challenge for the electrochemical deposition of iron chalcogenide. However, no one has succeeded to obtain superconducting FeSe films.[15-19] Here, we show a synthesis of superconducting FeSe films using the electrochemical method.

Electrochemical depositions were performed by a three-electrode method. Cathode, anode and reference electrode were Pt plate, Fe plate and Ag/AgCl electrode, respectively. In order to prepare electrolyte, we dissolved 0.01 mol/l FeSO$_4$•7H$_2$O and

0.005 mol/l $SeO_2$ into distilled water. The $HNO_3$ was used as pH adjustor. We performed cyclic voltammetry (CV) in a range of the electric potential between -0.4 and -2.5 V. The films were characterized by x-ray diffraction with Cu-K$\alpha$ radiation using the $2\theta$-$\theta$ method. The actual composition of the films was investigated by the energy dispersive x-ray spectrometry (EDX). The temperature dependence of magnetization was measured using a superconducting quantum interface device (SQUID) magnetometer with an applied field of 5 Oe.

Figure 1 shows a cyclic voltammogram corresponding to the electric potential dependence of current density $J$. With decreasing the electric potential, $J$ value decrease monotonically, and exhibit an anomaly around -1.75 V. In the reverse scan, $J$ value linearly increases. Assuming that this anomaly is indicative of the deposition of Fe-Se compounds, we performed the electrochemical synthesis in a potential range between -1.00 and -2.50 V in order to deposit superconducting FeSe films.

The electric potential dependence of the composition ratio of Fe and Se estimated by EDX measurements is shown in Fig. 2. At the electric potential of -1.00 V, Se was dominantly detected. At the electric potential between -1.50 and -2.50 V, Se and Fe composition ratio was estimated to be around 60 and 40 %. We found that the Se composition is slightly higher than that of Fe at the electric potentials between -1.50 and

-2.50 V.

The XRD patterns for the samples obtained at the electric potential of -1.00 and -2.50 V are summarized in Fig. 3. At the electric potential of -1.0 V, only hexagonal Se peaks were observed, which is consistent with EDX results. In contrast, we found that peaks identified as tetragonal FeSe were appeared in the samples at -1.5, -1.75 and -2.00 V. However, the peaks are noticeably broadened because of Se-rich composition. Since FeSe phase is stable in the slightly Fe-rich condition[4, 20], we tried to control the Fe/Se ratio by tuning of pH value.

Figure 4 shows the pH dependence of the composition ratio of Fe and Se analyzed by EDX measurements. Fe-rich condition was obtained in the range of pH 2.6 and 2.1. Figure 5 presents XRD patterns for FeSe films electrodeposited between pH 2.9 and 2.1. We have clearly observed sharp peaks corresponding to tetragonal FeSe, and the impurity phase of hexagonal FeSe was almost disappeared. The electric potential -1.75 V and pH 2.3 were found to be the optimal condition to electrodeposite tetragonal FeSe films. Figure 6 presents temperature dependence of magnetization of FeSe films synthesized at -1.75 V and pH 2.3. The data were obtained after field cool condition with magnetic field of 5 Oe. The magnetization suddenly decreased around 3.5 K, corresponding to superconducting transition.

In summary, we successfully synthesized superconducting FeSe films using the electrochemical method. Tetragonal FeSe films electrodeposited at -1.75 V and pH 2.3 exhibited superconducting transition $T_c$ ~ 3.5 K. Our results give the novel synthesis method to fabricate the superconducting coated conductor including films, wires tapes and so on.

**Acknowledgement**

This work was partly supported by a Grant-in-Aid for Scientific Research (KAKENHI).

**Figure caption**

Fig. 1 Cyclic voltammemtry (CV) measurement for distilled water dissolving 0.01 mol/l FeSO$_4$ ·7H$_2$O and 0.005 mol/l SeO$_2$.

Fig. 2 The electric potential dependence of the composition ratio of Fe and Se measured by EDX measurements.

Fig. 3 X-ray diffraction patterns of FeSe films electrodeposited by different electric potentials. Peaks marked by * and ▲ indicate hexagonal FeSe and Se, respectively.

Fig. 4 The pH value dependence of the composition ratio of Fe and Se measured by EDX measurements.

Fig. 5 X-ray diffraction patterns of FeSe thin films electrodeposited by the difference pH value. Peaks marked by * indicate hexagonal FeSe.

Fig. 6 Temperature dependence of magnetization for FeSe films electrodeposited at electric potential -1.75 V and pH 2.3.

**Reference**

[1] Y. Kamihara, T. Watanabe, M. Hirano, and H. Hosono, J. Am. Chem. Soc. **130**, 3296 (2008).

[2] M. Rotter, M. Tegel, and D. Johrendt, Phys. Rev. Lett. **101**, 107006 (2008).

[3] X. C. Wang, Q. Q. Liu, Y. X. Lv, W. B. Gao, L. X. Yang, R. C. Yu, F. Y. Li, and C. Q. Jin, Solid State Commun. **148**, 538 (2008).

[4] F. C. Hsu, J. Y. Luo, K. W. The, T. K. Chen, T. W. Huang, P. M. Wu, Y. C. Lee, Y. L. Huang, Y. Y. Chu, D. C. Yan, and M. K. Wu, Proc. Natl. Acad. Sci. U. S. A. **105**, 14262 (2008).

[5] H. Ogino, Y. Matsumura, Y. Katsura, K. Ushiyama, H. Horii, K. Kishio, and J. Shimoyama, Supercond. Sci. Technol. **22**, 075008 (2009).

[6] K. Ishida, Y. Nakai, and H. Hosono, J. Phys. Soc. Jpn. **78**, 062001 (2009).

[7] Y. Mizuguchi and Y. Takano, J. Phys. Soc. Jpn. **79**, 102001 (2010).

[8] Y. Mizuguchi, F. Tomioka, S. Tsuda, T. Yamaguchi, and Y. Takano, Appl. Phys. Lett. **93**, 152505 (2008).

[9] S. Margadonna, Y. Takabayashi, Y. Ohishi, Y. Mizuguchi, Y. Takano, T. Kagayama, T. Nakagawa, M. Takata, and K. Prassides, Phys. Rev. B **80**, 064506 (2009).

[10] S. Medvedev, T. M. McQueen, I. A. Troyan, T. Palasyuk, M. I. Eremets, R. J. Cava, S. Naghavi, F. Casper, V. Ksenofontov, G. Wortmann, and C. Felser, Nature Mater. **8**, 630 (2009).

[11] S. Masaki, H. Kotegawa, Y. Hara, H. Tou, K. Murata, Y. Mizuguchi, and Y. Takano, J. Phys. Soc. Jpn. **78**, 063704 (2009).

[12] M. J. Wang, J. Y. Luo, T. W. Hung, H. H. Chang, T. K. Chen, F. C. Hsu, C. T. Wu, P. M. Wu, A. M. Chang, and M. K. Wu, Phys. Rev. Lett. **103**, 117002 (2009).

[13] A. Agatsuma, T. Yamagishi, S. Takeda and M. Naito, Physica C **470**, 1468 (2010).

[14] H. Abe and K. Yoshii, Jpn. J. Appl. Phys. **41**, 685 (2002).

[15] Y. Mizuguchi, Y. Hara, K. Deguchi, S. Tsuda, T. Yamaguchi, K. Takeya, H. Kotegawa, H. Tou, and Y. Takano, Supercond. Sci. Technol. **23**, 054013 (2010).

[16] S. M. Pawer, A. V. Moholkar, U. B. Suryavanshi, K. Y. Rajpure, and C. H. Bhosale, Sol. Energy Mater. Sol. Cells **91**, 560 (2007).

[17] S. Thanikaikarasan, T. Mahalingam, K. Sundaram, A. Kathalingam, Yong Deak Kim, and Taekyu Kim, Vacuum **83**, 1066 (2009).

[18] S. Thanikaikarasan, T. Mahalingam, M. Raja, Taekyu Kim, and Yong Deak Kim, J. Mater Sci.: Mater Electron **20**, 727 (2009).

[19] P. Y. Chen, S. F. Hu, R. S. Liu, and C. Y. Huang, Thin Solid Films **519**, 8397 (2011).

[20] T. M. McQueen, Q. Huang, V. Ksenofontov, C. Felser, Q. Xu, H. Zandbergen, Y. S. Hor, J. Allred, A. J. Williams, D. Qu, J. Checkelsky, N. P. Ong, and R. J. Cava, Phys. Rev. B **79**, 014552 (2009).

Fig.1

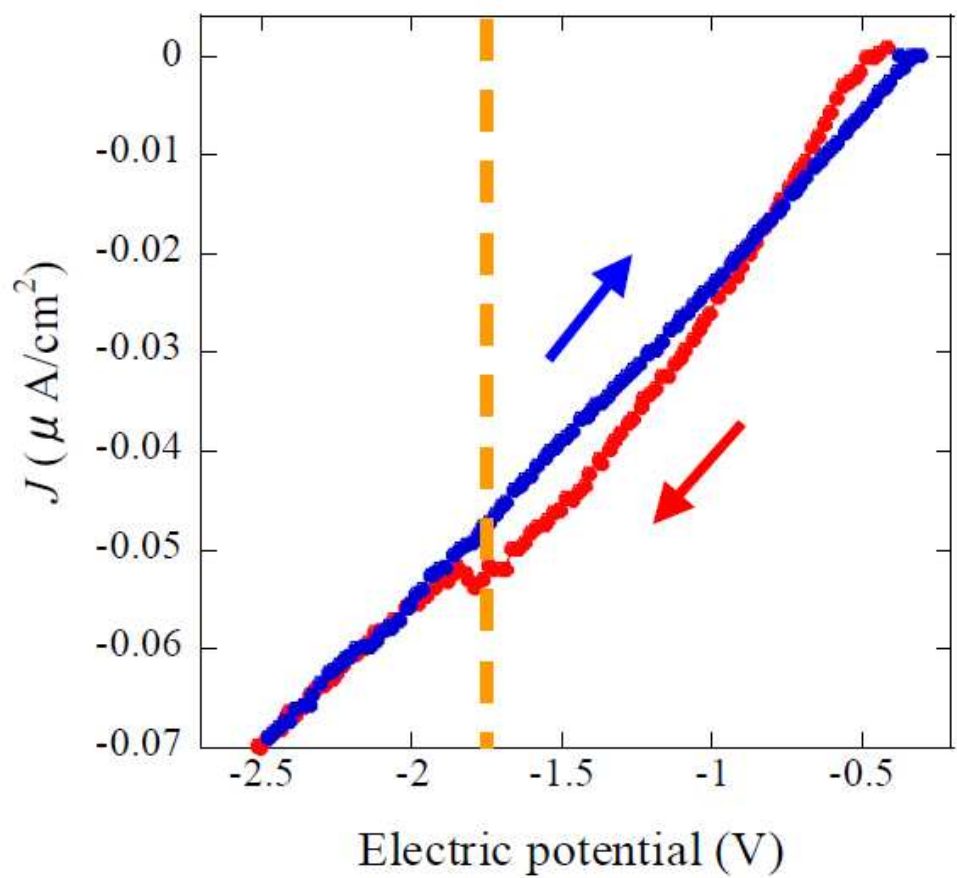

Fig.2

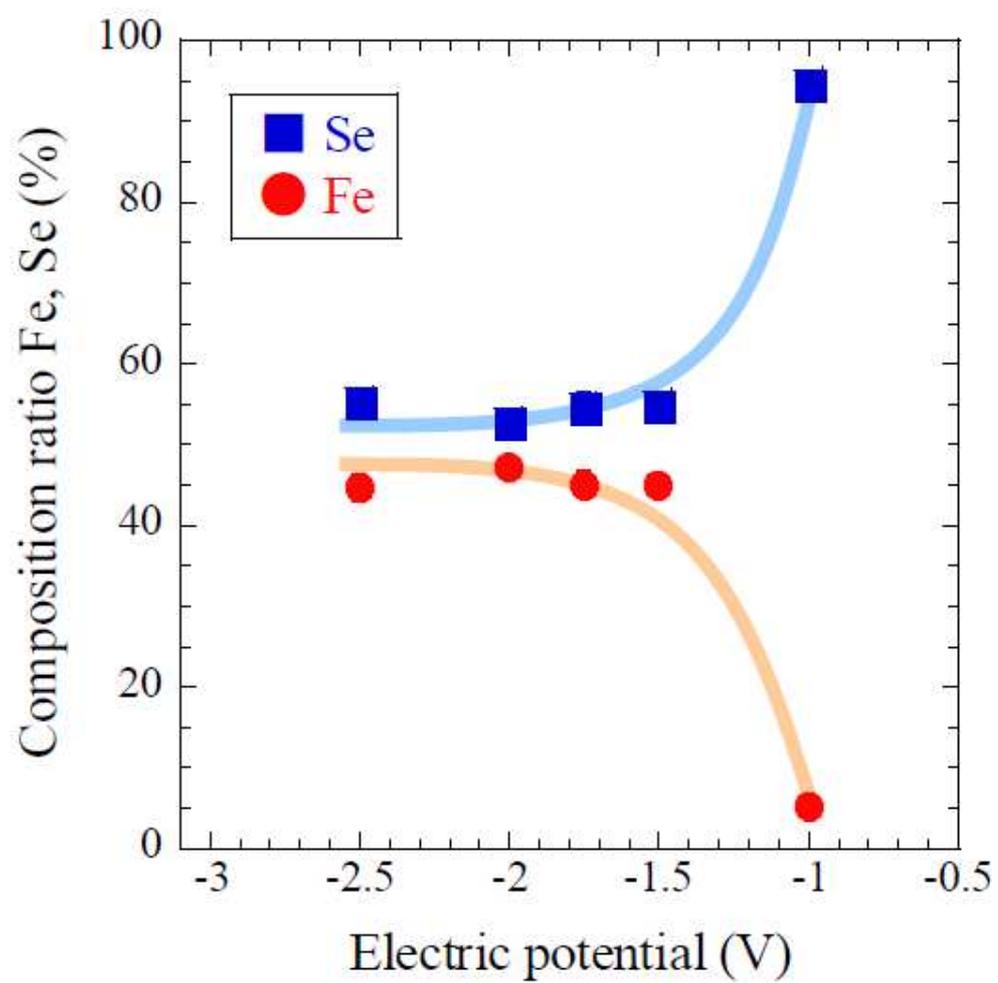

Fig. 3

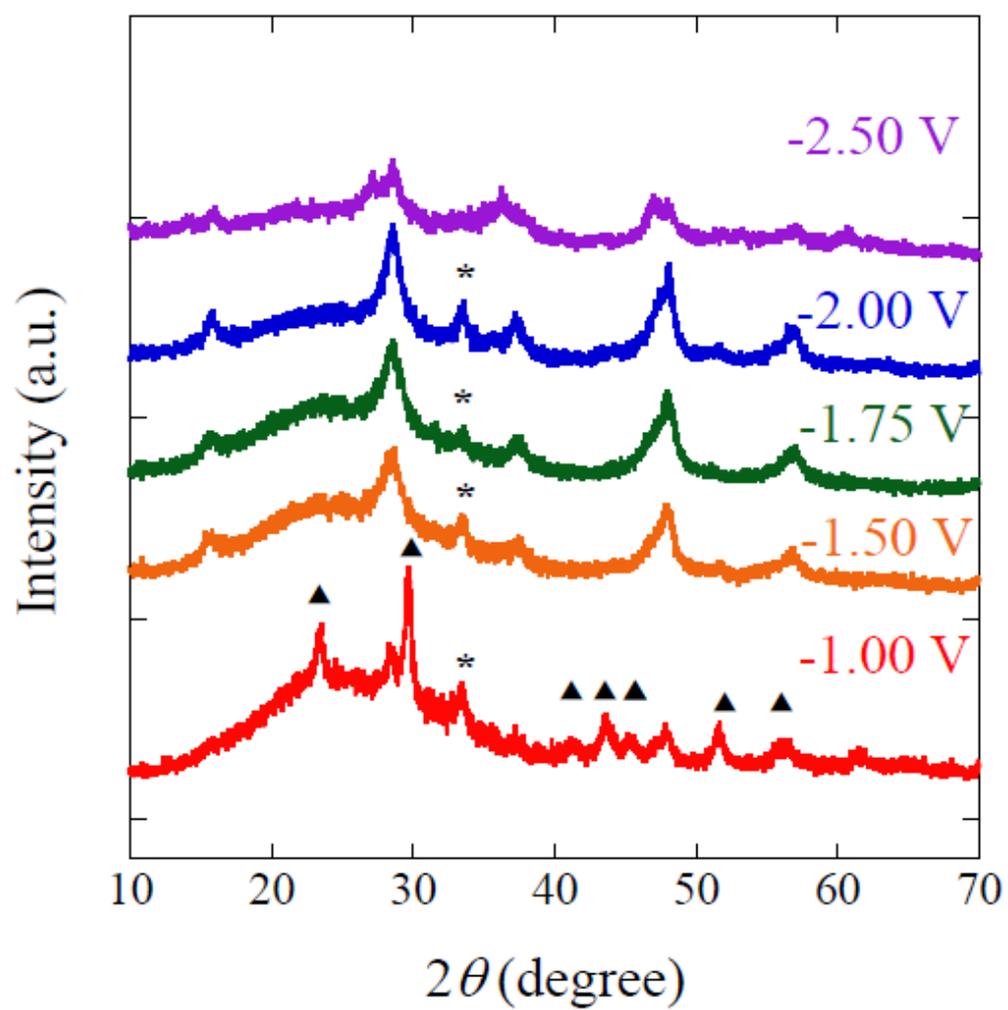

Fig.4

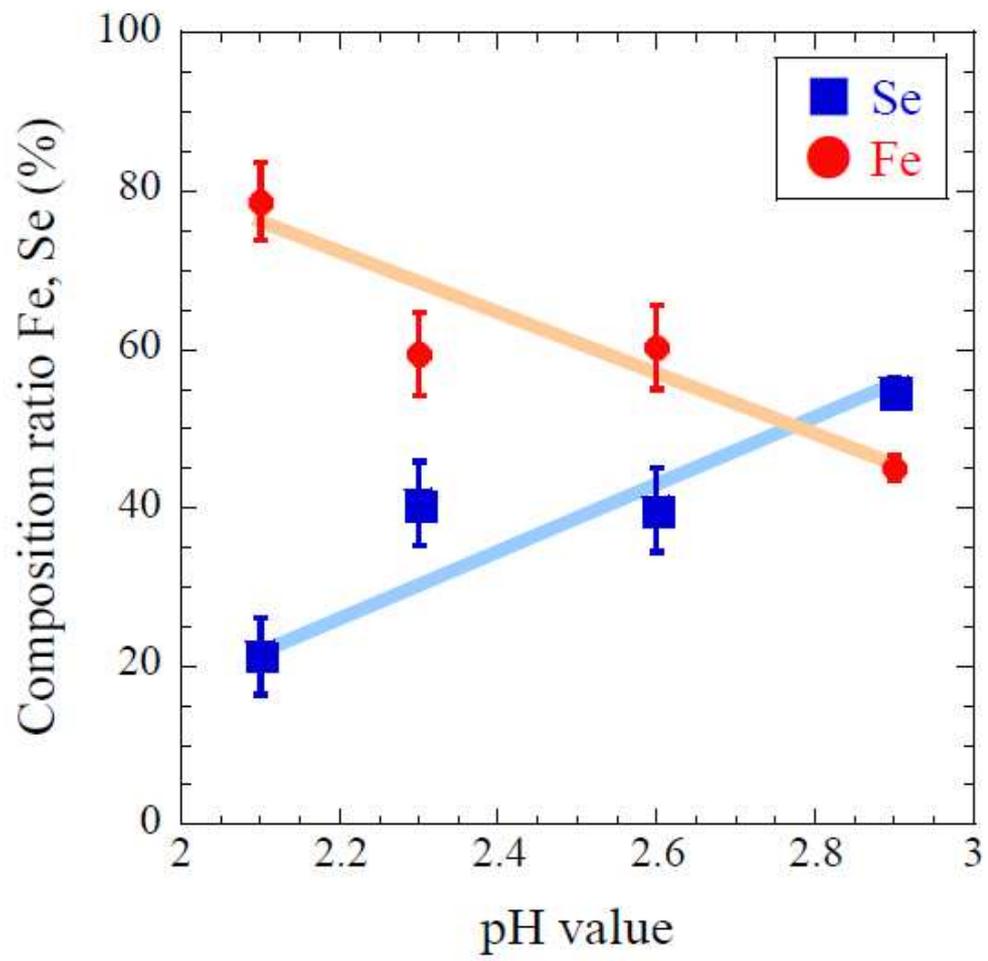

Fig. 5

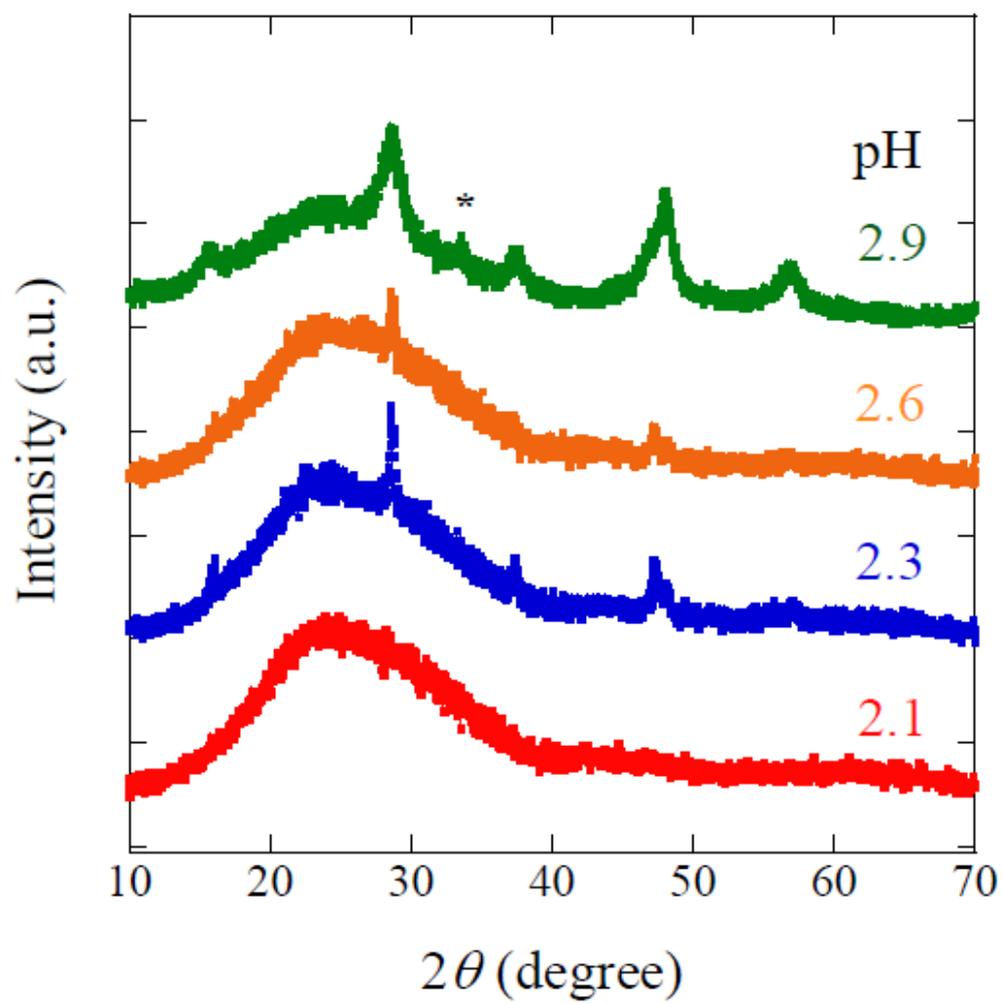

Fig. 6

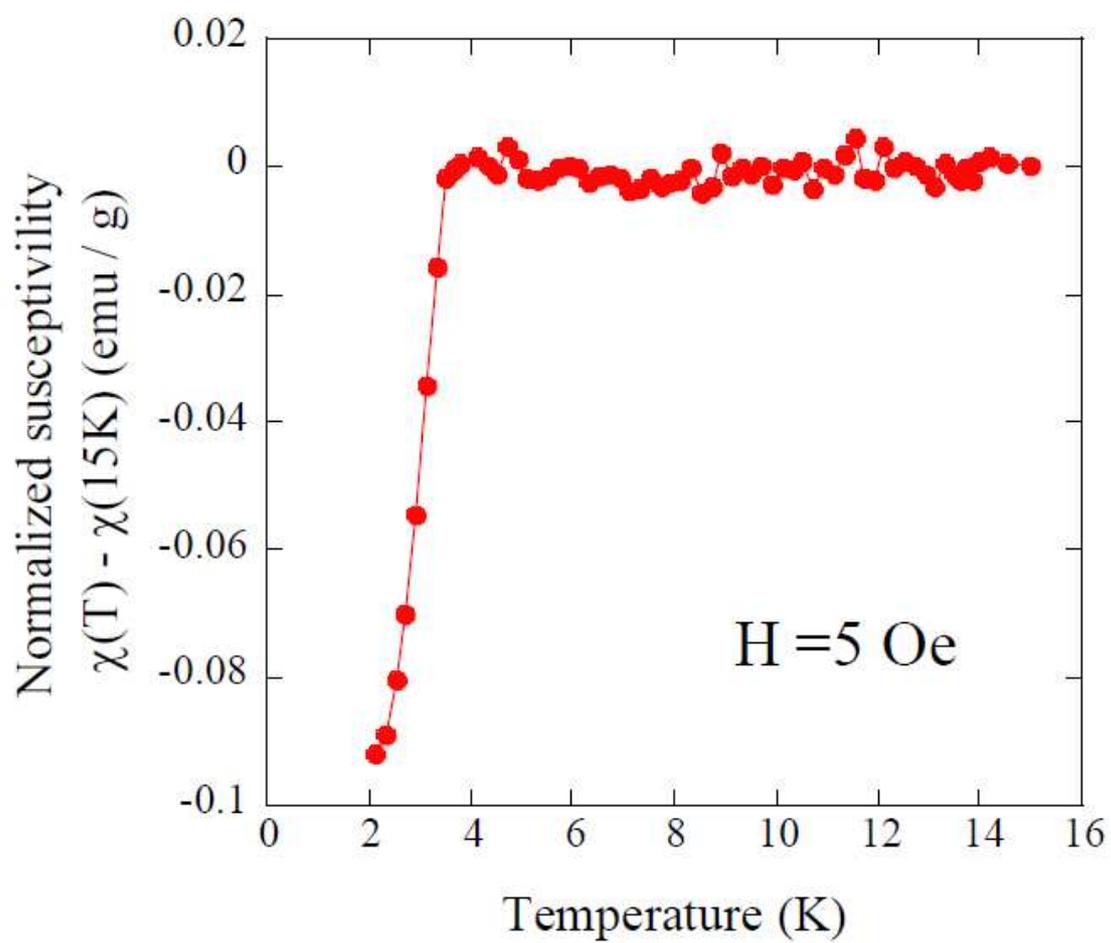